# Facile Algebraic Representation of a Novel Quaternary Logic

Ifat Jahangir[1*], Anindya Das[2], Masud Hasan[3]

[1]Department of Electrical Engineering, University of South Carolina, Columbia, SC 29208
[2]Department of Computer Science, Iowa State University, Ames, IA 50010
[3]Department of Computer Science, Taibah University, Madinah Munawarah, Saudi Arabia 41411
[*]Corresponding author, email: ifat00@gmail.com

*Abstract*—In this work, a novel quaternary algebra has been proposed that can be used to implement an arbitrary quaternary logic function in more than one systematic ways. The proposed logic has evolved from and is closely related to the Boolean algebra for binary domain; yet it does not lack the benefits of a higher-radix system. It offers seamless integration of the binary logic functions and expressions through a set of transforms and allows any binary logic simplification technique to be applied in quaternary domain. Since physical realization of the operators defined in this logic has recently been reported, it has become very important to have a well-defined algebra that will facilitate the algebraic manipulation of the novel quaternary logic and aid in designing various complex logic circuits. Therefore, based on our earlier works, here we describe the complete algebraic representation of this logic for the first time. The efficacy of the logic has been shown by designing and comparing several common logic circuits with existing designs in both binary and quaternary domain.

## I. INTRODUCTION

For many years digital devices have been designed using binary logic. Even today, the latest computing systems are designed and developed using only the binary logic. Since multi-valued logic enables more information to be packed in a single digit, researchers have been working on multi-valued logic for many years [1]-[18]. With the development of novel electronic and optical devices, it is now possible to implement circuits for more complicated logic systems [4]-[8]. Many of these devices are capable of dealing with more than two logic states, so their efficiency could be utilized if we use multi-valued logic for digital circuits. Some multi-valued logic systems such as ternary and quaternary logic schemes have been developed and they have been being experimented for a long time [1]. These logic systems are derived as propositional or quantum logic [1],[9].

Quaternary logic has several advantages over binary logic. Since it requires half the number of digits to store any information than its binary equivalent, it is good for storage; given that the quaternary storage mechanism is less than twice as complex as the binary system. For the same reason, quaternary devices require simpler parallel circuits to process same amount of data than that needed in binary logic devices. Inspired by such advantages, many researchers proposed different variants of quaternary logic in the past decades, demonstrated theoretically and experimentally [9]-[19].

Although there are numerous references on quaternary logic in the literature, we introduced yet another new and unique variant of quaternary logic for the first time in our earlier works [20]-[26]. This logic offers all the benefits of a higher radix system, yet can readily take advantage of existing binary circuit designs and design optimization rules which were developed over many decades of relentless effort by countless researchers. The simplicity and easy scalability of the common logic circuits offered by the new logic was evident in our earlier reports, where we presented the design of several types of adders, comparators, encoders and decoders [21]-[26]. As a matter of fact, it is possible to implement any quaternary function in two types of sum-of-products (SOP) expressions, one of them is only possible using the proposed quaternary logic. This SOP expression integrates existing designs and design methodologies in a systematic way, which can be optimized further through algebraic manipulation [25]. The novelty of this logic has drawn attention of many researchers working in the field of quaternary logic and as a result, very recently, there have been several reports on physical realization of this logic [27]-[30]. None of these works discuss the prospects and completeness of this logic as an extension of Boolean algebra, neither do we see a set of rules to facilitate the design of arbitrary functions that would meet the growing need of a general-purpose higher-radix logic system. Therefore, based on our earlier works, here we describe the complete algebraic representation of this logic for the first time. We use the electronic realization scheme demonstrated in [27]-[29] to calculate some physical parameters such as transistor count and gate depth in a logic circuit.

In our present work, we start our discussion in Section II with a formal description of the quaternary logic, including the definition and classification of the operators. Here we also briefly discuss the physical realization of the logic gates. In Section III, the fundamental properties of quaternary algebra and its operators are presented along with some important theorems. Then we present the method of expressing arbitrary quaternary functions in Section IV where two different representations of sum-of-products (SOP) expressions are shown. Section V is dedicated to the computation of theoretical upper bounds of gate count and gate depth for both forms of SOP discussed in Section IV. In Section VI, design of several combinational logic blocks such as multiplexer, decoder and demultiplexer are shown using the proposed quaternary algebra.

Based on these designs, we present a comparative analysis of the different variants of quaternary logic in Section VII.

## II. QUATERNARY ALGEBRA

Quaternary algebra is defined as a set of operators and a set of values {0, 1, 2, 3} for any valid proposition. Quaternary digits {0, 1, 2, 3} can be imagined as 2-bit binary equivalents 00, 01, 10, 11. A single quaternary digit is called a qudit when it is expressed as a number. If the bits of the binary equivalent of a qudit interchange their positions and still the quaternary state remains unchanged, then it is said to have binary symmetry; otherwise it is asymmetrical. It should be noted that quaternary states 0 and 3 are symmetrical, while 1 and 2 are asymmetrical.

### A. Classification of quaternary operators

There are several operators in the proposed quaternary algebra which are sufficient to describe any quaternary function. We classify these operators in two classes.

*1) Fundamental Operators:*

Fundamental operators are those selected operators that are sufficient to completely define the quaternary algebra and can be used to derive other operators.

*2) Functional operators:*

The functional operators are those operators that can be expressed by a combination of two or more fundamental operators.

It will be shown later that functional operators can also be used to express any arbitrary quaternary function; the reason behind this classification lies in our consideration of generality and flexibility. In subsection III-C, we will show three sets of operators, comprising both fundamental and functional operators, each set being sufficient for expressing any arbitrary quaternary function. This redundancy is allowed in the logic system for practical purposes - each set offers certain distinct benefits when it comes to physical realization, yet all of them are connected through various laws of the algebra. Therefore, the operators offering the most flexible and wide range of applications are chosen as fundamental operators, and the rest are defined as their derivatives (functional operators). This will be discussed in more details in subsequent sections.

TABLE I. CLASSIFICATION OF QUATERNARY OPERATORS

| Quaternary Operators | |
|---|---|
| Fundamental Operators | Functional Operators |
| AND, OR, NOT, Bitswap | Inward Inverter, Outward Inverter, Equality, MIN, MAX, XOR |

### B. Definition of quaternary operators

A quaternary digit can be expressed by two binary digits packed together using the following notion -

$$A = \langle a_1, a_0 \rangle \equiv (2 \times a_1 + a_0)_{10} \qquad (1)$$

where $a_1$ and $a_0$ are the constituent bits of the quaternary digit $A$ and the right side of (1) denotes the magnitude of $A$ in decimal system. In general, the fundamental dyadic operators work like bitwise binary operators if the above notion is adopted -

$$F(A, B) = F(\langle a_1, a_0 \rangle, \langle b_1, b_0 \rangle) = \langle f(a_1, b_1), f(a_0, b_0) \rangle \qquad (2)$$

where $F$ and $f$ stands for similar quaternary and binary operators respectively. The above notation of expressing quaternary digits (operators) in terms of binary digits (operators) is called packed-binary representation of quaternary digits (operators).

The mathematical symbols and truth tables of all operators are shown in Table II. In Table II, some monadic/unary operators have different symbols from our earlier works [20]-[24] to improve readability and facilitate type-setting; the symbol of outward inverter $\hat{A}$ is changed to $!A$ and the overhead symbol of bitswap $\tilde{A}$ is changed to $\sim A$.

Bitswap is the only fundamental operator that does not have any binary equivalent and is unique in this algebra (first presented in [24]); it swaps the two bits of the binary equivalent of the quaternary operand. It leaves the symmetrical numbers unchanged but inverts (i.e. NOT) the asymmetrical numbers, so this operator can also be defined in the following way-

$$\text{Binary Bitswap}, \sim a = \begin{cases} \bar{a} & ; \quad a \text{ asymmetric} \\ a & ; \quad a \text{ symmetric} \end{cases} \qquad (3)$$

Using packed-binary representation, the NOT operator can be expressed in the following way -

$$\overline{A} = \overline{\langle a_1, a_0 \rangle} = \langle \overline{a_1}, \overline{a_0} \rangle \qquad (4)$$

On the other hand, bitswap can be expressed as

$$\sim A = \sim \langle a_1, a_0 \rangle = \langle a_0, a_1 \rangle \qquad (5)$$

When the bitswap operator follows another operator, we get a compound form of operators that may be realizable directly depending on the technology. Some examples are bitswap AND (AND followed by bitswap), bitswap NOR (NOR followed by bitswap), bitswap XNOR (XNOR followed by bitswap), etc. In the bitswap NAND, NOR, NOT and XNOR, the inverter is obviously "NOT", not the inward or outward inverter. However, if an outward or inward inverter follows another operator, that is clearly mentioned, such as outward AND, inward XOR, etc. Fig. 1 shows the circuit symbols of all the fundamental and functional operators.

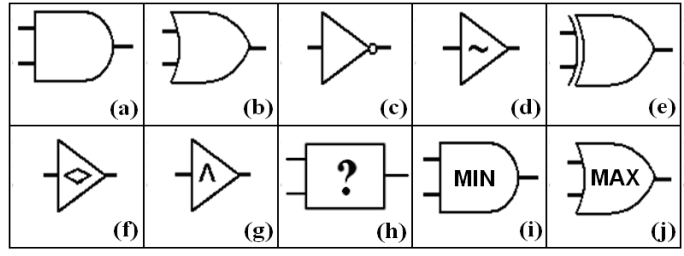

Fig. 1. Circuit symbols of quaternary operators: (a) AND, (b) OR, (c) NOT, (d) Bitswap, (e) XOR, (f) Inward Inverter, (g) Outward Inverter, (h) Equality, (i) MIN, (j) MAX.

TABLE. II: MATHEMATICAL SYMBOLS AND TRUTH TABLES OF QUATERNARY OPERATORS [†]

| Operands | $A$ | 0 | 0 | 0 | 0 | 1 | 1 | 1 | 2 | 2 | 3 |
|---|---|---|---|---|---|---|---|---|---|---|---|
|  | $B$ | 0 | 1 | 2 | 3 | 1 | 2 | 3 | 2 | 3 | 3 |
| NOT | $\overline{A}$ | 3 | 3 | 3 | 3 | 2 | 2 | 2 | 1 | 1 | 0 |
| Outward inverter | $!A$ | 3 | 3 | 3 | 3 | 3 | 3 | 3 | 0 | 0 | 0 |
| Bitswap | $\sim A$ | 0 | 0 | 0 | 0 | 2 | 2 | 2 | 1 | 1 | 3 |
| Inward inverter | $A'$ | 2 | 2 | 2 | 2 | 2 | 2 | 2 | 1 | 1 | 1 |
| AND | $A \cdot B$ | 0 | 0 | 0 | 0 | 1 | 0 | 1 | 2 | 2 | 3 |
| OR | $A + B$ | 0 | 1 | 2 | 3 | 1 | 3 | 3 | 2 | 3 | 3 |
| XOR | $A \oplus B$ | 0 | 1 | 2 | 3 | 0 | 3 | 2 | 0 | 1 | 0 |
| Equality [‡] | $E(A,B)$ | 3 | 0 | 0 | 0 | 3 | 0 | 0 | 3 | 0 | 3 |
| MIN | $\underline{A \cdot B}$ | 0 | 0 | 0 | 0 | 1 | 1 | 1 | 2 | 2 | 3 |
| MAX | $\underline{A + B}$ | 0 | 1 | 2 | 3 | 1 | 2 | 3 | 2 | 3 | 3 |

[‡] Alternative symbol for equality used primarily in SOP expressions, $A^B = B^A = E(A, B)$.
[†] All dyadic operators are commutative with $F(A,B) = F(B,A)$. So identical pairs of $(A,B)$ are mentioned only once by showing 10 out of 16 possible combinations.

The equality operator is defined as -

$$E(A,B) = E(B,A) = A^B = B^A = \begin{cases} 0 & ; A \neq B \\ 3 & ; A = B \end{cases} \quad (6)$$

Using packed-binary representation, the functional inverters can be expressed as

$$!A = !\langle a_1, a_0 \rangle = \langle \overline{a_1}, \overline{a_1} \rangle \quad (7)$$

$$A' = \langle a_1, a_0 \rangle' = \langle \overline{a_1}, a_1 \rangle \quad (8)$$

## C. Required Sets of Operators and Their Physical Realization

There are three sets of operators in the proposed algebra, each of which is sufficient to express any quaternary function algebraically. These sets are listed below –
(1) AND, OR, NOT, bitswap
(2) AND, OR, equality
(3) MIN, MAX, equality

From the above list, the first two sets are used in form-II and form-I of sum-of-products (SOP) expressions respectively, which will be discussed in Section IV. The third set can also be used as an alternative representation of form-I of SOP as shown in [17], this depends on the choice of physical realization. Besides, MIN and MAX functions can be more efficient in sequential circuits if we compare the design in [17] with the ones in [20]. However, MIN (MAX) and AND (OR) are equivalent in the physical realization scheme assumed in this work, making them interchangeable if needed.

Another important reason for preferring the first set over the others is the fact that all operators in the first set have various properties that facilitate algebraic manipulation and simplification. De Morgan's law for the NOT operator, the distributive property of bitswap operator are two examples that are used widely to simplify many complex expressions. On the other hand, the equality operator is rather less flexible and we will show later that expressions containing the equality operator are often broken down in terms of the operators listed in the first set to facilitate simplification. Besides, the use of NOT and bitswap enables us to utilize the axioms of Boolean algebra and many existing techniques of binary logic design in the quaternary domain. For these reasons, the operators in the first set are chosen to be the fundamental operators and all other operators are described as their derivatives (please refer to Appendix A, where equality, MIN and MAX are expressed using AND, OR, NOT and bitswap).

Gogna et al. and Jain et al., in their recent works, reported multiple quantum well based spatial wavefunction-switched field effect transistors (SWSFET) to be suitable candidates for arbitrary quaternary operators [27]-[29]. Chattopadhyay et al. also proposed a polarization-based all-optical scheme for realizing the quaternary logic [30]. The design given in [27]-[29] performs a look-up table-based operation using multiple voltage lines connected to different quantum wells formed by heteroepitaxial superlattice structures. According to their design, any unary quaternary operator (inverters, bitswap, etc) can be realized by using just one SWSFET; for two-input operators at most five SWSFETs are required. However, for two-input OR, AND, MIN, MAX gates, only three SWSFETs are needed. For equality operator, five SWSFETs are required if both inputs are variable; however, only one SWSFET is needed if only one input is variable and the other is fixed, making it a unary operator.

## III. FUNDAMENTAL PROPERTIES OF QUATERNARY ALGEBRA AND ITS OPERATORS

In this section we will present the fundamental properties of quaternary algebra. Then some very important properties of quaternary operators will be discussed. These properties are helpful to express and manipulate complicated functions algebraically to ensure efficient implementation.

### A. Properties of Quaternary Algebra

The packed-binary representation of quaternary digits and operators show that all fundamental operators except the bitswap obey the axioms and properties of Boolean operators. Most of these properties have their dual forms, where AND and OR operators are interchanged, at the same time the constants are inverted via NOT. The properties given below show that our proposed logic satisfies all the requirements to be treated as algebra, as postulated by Huntington [31].

*1) Closure :*

For every dyadic operator, $F(A,B) \in \{0,1,2,3\}$, which is evident from definition. For every unary operator, $G(A) \in \{0,1,2,3\}$.

*2) Complement:*

There exists a unary operator NOT for which the following properties are true-

(1) $A + \overline{A} = \langle a_1 + \overline{a_1}, a_0 + \overline{a_0} \rangle = \langle 1,1 \rangle = 3$      (9)

(2) $A . \overline{A} = \langle a_1 . \overline{a_1}, a_0 . \overline{a_0} \rangle = \langle 0,0 \rangle = 0$      (10)

*3) Associativity:*

(1) $A + (B + C) = \langle a_1 + (b_1 + c_1), a_0 + (b_0 + c_0) \rangle = \langle (a_1 + b_1) + c_1, (a_0 + b_0) + c_0 \rangle = (A + B) + C$      (11)

(2) $A . (B . C) = \langle a_1 . (b_1 . c_1), a_0 . (b_0 . c_0) \rangle = \langle (a_1 . b_1) . c_1, (a_0 . b_0) . c_0 \rangle = (A . B) . C$      (12)

*4) Commutativity:*

(1) $A + B = \langle a_1 + b_1, a_0 + b_0 \rangle = \langle b_1 + a_1, b_0 + a_0 \rangle = B + A$      (13)

(2) $A . B = \langle a_1 . b_1, a_0 . b_0 \rangle = \langle b_1 . a_1, b_0 . a_0 \rangle = B . A$      (14)

*5) Distributivity:*

(1) $A + (B . C) = \langle a_1 + (b_1 . c_1), a_0 + (b_0 . c_0) \rangle = \langle (a_1 + b_1).(a_1 + c_1), (a_0 + b_0).(a_0 + c_0) \rangle = (A + B).(A + C)$      (15)

(2) $A . (B + C) = \langle a_1 . (b_1 + c_1), a_0 . (b_0 + c_0) \rangle = \langle (a_1.b_1) + (a_1.c_1), (a_0.b_0) + (a_0.c_0) \rangle = (A.B) + (A.C)$      (16)

*6) Boundedness:*

(1) $A + 0 = \langle a_1 + 0, a_0 + 0 \rangle = \langle a_1, a_0 \rangle = A$      (17)

$A . 3 = \langle a_1 . 1, a_0 . 1 \rangle = \langle a_1, a_0 \rangle = A$      (18)

(2) $A + 3 = \langle a_1 + 1, a_0 + 1 \rangle = \langle 1,1 \rangle = 3$      (19)

$A . 0 = \langle a_1 . 0, a_0 . 0 \rangle = \langle 0,0 \rangle = 0$      (20)

### B. Properties of Quaternary Operators

*1) Bitswap operator distributes itself over AND and OR operators.*

$\sim (A + B) = \sim \langle a_1 + b_1, a_0 + b_0 \rangle = \sim \langle a_1, a_0 \rangle + \sim \langle b_1, b_0 \rangle = \sim A + \sim B$      (21)

$\sim (A.B) = \sim (\langle a_1, a_0 \rangle . \langle b_1, b_0 \rangle) = \sim \langle a_1, a_0 \rangle . \sim \langle b_1, b_0 \rangle = \sim A . \sim B$      (22)

*2) NOT obeys the De Morgan's law, when applied to the output of OR or AND gates.*

$\overline{A + B} = \overline{\langle a_1 + b_1, a_0 + b_0 \rangle} = \langle \overline{a_1}.\overline{b_1}, \overline{a_0}.\overline{b_0} \rangle = \overline{A}.\overline{B}$      (23)

$\overline{A.B} = \overline{\langle a_1.b_1, a_0.b_0 \rangle} = \langle \overline{a_1} + \overline{b_1}, \overline{a_0} + \overline{b_0} \rangle = \overline{A} + \overline{B}$      (24)

*3) Like NOT, outward inverter also obeys the De Morgan's law, when applied to the output of OR or AND gates.*

$!(A + B) = !\langle a_1 + b_1, a_0 + b_0 \rangle = \langle \overline{a_1 + b_1}, \overline{a_1 + b_1} \rangle = \langle \overline{a_1}, \overline{a_1} \rangle . \langle \overline{b_1}, \overline{b_1} \rangle = !A . !B$      (25)

$$!(A.B) = !\langle a_1.b_1, a_0.b_0 \rangle = \langle \overline{a_1.b_1}, \overline{a_1.b_1} \rangle = \langle \overline{a_1}, \overline{a_1} \rangle + \langle \overline{b_1}, \overline{b_1} \rangle = !A + !B \tag{26}$$

4) There is no compact expression that can be used to express the distribution of inward inverter over AND or OR operators.

$$(A+B)' = (\langle a_1+b_1, a_0+b_0 \rangle)' = \langle \overline{a_1+b_1}, a_1+b_1 \rangle'$$

$$(A.B)' = (\langle a_1.b_1, a_0.b_0 \rangle)' = \langle \overline{a_1.b_1}, a_1.b_1 \rangle$$

None of the above can be expressed in a form similar to $\langle f(a_1,b_1), f(a_0,b_0) \rangle$. Thus, there is no algebraic expression to expand the operation of inward inverter following the AND or OR operation.

5) The order of inward inverter and NOT can be reversed.

$$(\overline{A})' = \langle \overline{a_1}, \overline{a_0} \rangle' = \langle \overline{a_1}, \overline{a_1} \rangle = \overline{\langle \overline{a_1}, a_1 \rangle} = \overline{(A')} \tag{27}$$

6) The order of outward inverter and NOT can be reversed.

$$!(\overline{A}) = !\langle \overline{a_1}, \overline{a_0} \rangle = \langle a_1, a_1 \rangle = \overline{\langle \overline{a_1}, \overline{a_1} \rangle} = \overline{(!A)} \tag{28}$$

7) The order of bitswap and NOT can be altered.

$$\sim(\overline{A}) = \sim \langle \overline{a_1}, \overline{a_0} \rangle = \langle \overline{a_0}, \overline{a_1} \rangle = !\langle a_0, a_1 \rangle = \overline{(\sim A)} \tag{29}$$

8) The order of bitswap and inward inverter can be altered under certain condition, not generally.

$$(\sim A)' = (\langle a_0, a_1 \rangle)' = \langle \overline{a_0}, a_0 \rangle \tag{30}$$

$$\sim(A') = \sim \langle \overline{a_1}, a_1 \rangle = \langle a_1, \overline{a_1} \rangle \tag{31}$$

This implies, $(\sim A)' = \sim(A')$ if and only if $a_1 = \overline{a_0}$, i.e. $A$ is asymmetric.

9) The order of bitswap and outward inverter can be altered under certain condition, not generally.

$$\sim(!A) = \sim \langle \overline{a_1}, \overline{a_1} \rangle = \langle \overline{a_1}, \overline{a_1} \rangle \tag{32}$$

$$!(\sim A) = !\langle a_0, a_1 \rangle = \langle \overline{a_0}, \overline{a_0} \rangle \tag{33}$$

This implies, $\sim(!A) = !(\sim A)$ if and only if $a_1 = a_0$, i.e. $A$ is symmetric.

10) The order of inward and outward inverters can never be reversed under any condition.

$$(!A)' = (\langle \overline{a_1}, \overline{a_1} \rangle)' = \langle a_1, \overline{a_1} \rangle \tag{34}$$

$$!(A') = !\langle \overline{a_1}, a_1 \rangle = \langle a_1, a_1 \rangle \tag{35}$$

This implies, $(!A)' \neq !(A')$ under any circumstances.

### C. Theorems of Quaternary Algebra:

There are several theorems in the proposed quaternary algebra that are derived from the fundamental postulates of the algebra and properties of the operators. Here we present a list of theorems that are useful in algebraic operations -

1) The Law of Idempotency:

$$A + A = A, \quad A.A = A \tag{36}$$

2) The Law of Absorption:

$$A + (A.B) = A, \quad A.(A+B) = A \tag{37}$$

3) The Law of Identity:

$$A + B = A, \quad A.B = A \quad ; \text{ for } A = B \tag{38}$$

4) The Law of Complements with NOT:

$$A + B = 3, \quad A.B = 0 \quad ; \text{ for } A = \overline{B} \tag{39}$$

5) The Law of Involution with NOT and bitswap:

$$A = \overline{\overline{A}}, \quad A = \sim(\sim A) \tag{40}$$

*6) The Law of Elimination with NOT:*

$$X+\overline{X}.Y = X + Y, \quad X.(\overline{X}+Y) = X.Y \tag{41}$$

*7) The Law of Concensus with NOT:*

$$X.Y+\overline{X}.Z+Y.Z = X.Y + \overline{X}.Z \tag{42}$$

$$(X+Y).(\overline{X}+Z).(Y+Z) = (X+Y).(\overline{X}+Z) \tag{43}$$

*8) The Law of Interchange with NOT:*

$$(X.Y)+(\overline{X}.Z) = (X+Y).(\overline{X}+Z) \tag{44}$$

$$(X+Y).(\overline{X}+Z) = (X.Z)+(\overline{X}.Y) \tag{45}$$

## IV. EXPRESSION OF ARBITRARY FUNCTIONS IN QUATERNARY ALGEBRA

In this section, we will show the completeness of our proposed logic algebra by demonstrating that any arbitrary quaternary function can be expressed in terms of the operators described in Section II. We will demonstrate two forms of SOP (sum-of-products) to express any function.

To describe a set of quaternary variables, we will often use the array notation. For example, if a function $F$ takes $n$ inputs namely $X_1, X_2, X_3, \ldots X_n$ and gives a single output, then we write the variables in array form as $\mathbf{X} = \{X_1, X_2, X_3, \ldots, X_n\}$ and the function is written as $F(\mathbf{X})$. Similarly $F(\mathbf{X},\mathbf{Y})$ takes two such array operands of same length and gives a single scalar output. Like functions, operators can also handle arrays. For example, single-output OR and AND operators with array inputs are given below:

$$\sum \mathbf{X} \equiv \sum_{i=1}^{n} X_i \equiv X_1 + X_2 + \ldots + X_n \tag{46}$$

$$\prod \mathbf{X} \equiv \prod_{i=1}^{n} X_i \equiv X_1.X_2 \ldots X_n \tag{47}$$

However, functions (operators) with multiple parallel instances can be expressed as function (operator) arrays. For functions arrays, both inputs and functions are identified in boldface. The functions (operators) take one or more input arrays of same length and generate an output array with same length. Some examples are given below-

$$\mathbf{F}(\mathbf{X},\mathbf{Y}) \equiv \{F(X_1,Y_1), \ldots F(X_n,Y_n)\} \tag{48}$$

$$\mathbf{E}(\mathbf{A},\mathbf{B}) = \begin{cases} 0 & ; \text{ for every } A_i \neq B_i \\ 3 & ; \text{ for every } A_i = B_i \end{cases} \tag{49}$$

$$\overline{\mathbf{X}} \equiv \{\overline{X_1}, \overline{X_2}, \ldots, \overline{X_n}\} \tag{50}$$

$$\mathbf{X}+\mathbf{Y} \equiv \{X_1+Y_1, X_2+Y_2, \ldots X_n+Y_n\} \tag{51}$$

$$\sum \mathbf{X}.\mathbf{Y} \equiv X_1.Y_1 + X_2.Y_2 + \ldots + X_n.Y_n \tag{52}$$

### A. Implementation of any function in quaternary algebra (Form-I of SOP):

**Lemma 1:** It is possible to generate a minterm with any value i.e. 1, 2 or 3 for a particular set of input values for a finite number of variables using only the equality and AND operators.

Let us consider a set of $n$ variables $\mathbf{X}=\{X_1, X_2, \ldots, X_n\}$. For a particular set of inputs $\mathbf{V} = \{V_1, V_2, \ldots, V_n\}$ where $V_i \in \{0,1,2,3\}$, a function $M_D(\mathbf{X},\mathbf{V})$ would produce an output of $0$ or $D$, where $D$ is 1, 2 or 3. The functions defined as follows:

$$M_D(\mathbf{X},\mathbf{V}) = \begin{cases} D & ; \text{ if } \mathbf{X} = \mathbf{V} \\ 0 & ; \text{ otherwise} \end{cases} \tag{53}$$

If $D = 3$, we can define a function $G(\mathbf{X},\mathbf{V})$ as follows:

$$G(\mathbf{X},\mathbf{V}) = \begin{cases} 3 & ; \text{ if } \mathbf{X} = \mathbf{V} \\ 0 & ; \text{ otherwise} \end{cases} = \prod \mathbf{E}(\mathbf{X},\mathbf{V}) \tag{54}$$

If $\mathbf{X}$ and $\mathbf{V}$ are equal, only then we get $G(\mathbf{X},\mathbf{V}) = 3$.
From (53) and (54), we can write

$$M_D(\mathbf{X},\mathbf{V}) = \left[\prod \mathbf{E}(\mathbf{X},\mathbf{V})\right].D \tag{55}$$

We call $M_D(\mathbf{X},\mathbf{V})$ a minterm for quaternary algebra with output $D$. If we set $D = 3$, then from (55), $M_3(\mathbf{X},\mathbf{V}) = G(\mathbf{X},\mathbf{V})$. In this derivation, both $\mathbf{X}$ and $\mathbf{V}$ are taken arbitrarily, so we can say that any minterm can be expressed with only the equality and AND operators which can generate a desired output value (1, 2 or 3) for a defined set of input values for a finite number of variables.

**Lemma 2**: It is possible to implement any function using only equality, AND and OR operators.

Let us implement a function of $n$ input variables, the set of which is given by $\mathbf{X}$. For each combination of inputs $\mathbf{V}$, there is a minterm $M_D(\mathbf{X},\mathbf{V})$ with output value $D$. Since $\mathbf{X}$ can match with at most one input set $\mathbf{V}$, only one minterm can produce a non-zero output value. Thus, using Lemma 1, we can express the function in the following form:

$$F(\mathbf{X}) = \sum M_1(\mathbf{X},\mathbf{V}_1) + \sum M_2(\mathbf{X},\mathbf{V}_2) + \sum M_3(\mathbf{X},\mathbf{V}_3)$$
$$= \sum \left[\prod E(\mathbf{X},\mathbf{V}_1)\right].1 + \sum \left[\prod E(\mathbf{X},\mathbf{V}_2)\right].2 + \sum \left[\prod E(\mathbf{X},\mathbf{V}_3)\right].3 \qquad (56)$$

where $\mathbf{V}_1$, $\mathbf{V}_2$ and $\mathbf{V}_3$ are sets of input combinations for which the minterms will produce outputs of 1, 2 and 3, respectively. For all other input combinations, the function returns 0 as an output. Eq. (56) defines an expression for *sum-of-product* (SOP) that can be used to implement arbitrary functions. We call it form-I of SOP. An example of form-I is given below by defining an arbitrary function -

$$F(X,Y,Z) = \begin{cases} 1; & \text{for } \{X,Y,Z\} = \{1,2,0\} \\ 2; & \text{for } \{X,Y,Z\} = \{3,1,2\} \\ 3; & \text{for } \{X,Y,Z\} = \{2,3,1\} \end{cases} \qquad (57)$$

Therefore, according to (56), we can write $F(X,Y,Z)$ as follows:
$$F(X,Y,Z) = X^1.Y^2.Z^0.1 + X^3.Y^1.Z^2.2 + X^2.Y^3.Z^1.3 \qquad (58)$$

### B. Development of Form-II of SOP

From the definition of OR, we know $1 + 2 = 3$. We can use it in (55) to decompose $M_3$ in two components:
$$M_3(\mathbf{X},\mathbf{V}) = \left[\prod E(\mathbf{X},\mathbf{V})\right].3 = \left[\prod E(\mathbf{X},\mathbf{V})\right].(1+2) = M_1(\mathbf{X},\mathbf{V}) + M_2(\mathbf{X},\mathbf{V}) \qquad (59)$$

Now, if we have $k_1$ minterms with output 1, $k_2$ minterms with output 2 and $k_3$ minterms with output 3, then using the decomposition in (59), we can write

$$M_i = \begin{cases} M_1; & \text{for } k_1 + k_3 \text{ terms} \\ M_2; & \text{for } k_2 + k_3 \text{ terms} \end{cases} \qquad (60)$$

Let us write $M_i(\mathbf{X},\mathbf{V})$ and $G(\mathbf{X},\mathbf{V})$ in packed-binary form:
$$M_i(\mathbf{X},\mathbf{V}) = \langle m_{i1}, m_{i0} \rangle \qquad (61)$$
$$G(\mathbf{X},\mathbf{V}) = \langle g_1, g_0 \rangle \qquad (62)$$

From (53)-(55), we get the following relation:
$$M_i(\mathbf{X},\mathbf{V}) = \langle g_1, g_0 \rangle.i = \begin{cases} \langle g_1, g_0 \rangle.\langle 0,1 \rangle & ; i = 1 \\ \langle g_1, g_0 \rangle.\langle 1,0 \rangle & ; i = 2 \end{cases} \qquad (63)$$

After simplification, $M_1$ and $M_2$ can be written as
$$M_1(\mathbf{X},\mathbf{V}) = \langle 0, g_0 \rangle \qquad (64)$$
$$M_2(\mathbf{X},\mathbf{V}) = \langle g_1, 0 \rangle \qquad (65)$$

Using bitswap on both sides of (65),
$$\sim M_2(\mathbf{X},\mathbf{V}) = \langle 0, g_1 \rangle \qquad (66)$$

Since both $g_0$ and $g_1$ are dyadic functions, we have effectively converted $M_1$ and $M_2$ into binary equivalent functions.

Now, both $g_0$ and $g_1$ are functions of $n$ quaternary variables, which are equivalent to $2n$ binary variables written in packed-binary form. In a quaternary SOP, we have multiple minterms; but all minterms will have the form of either $M_1$ or $M_2$. Using (64) and (66), we can get the binary equivalents of all such minterms and vice versa. Assuming all inputs and outputs to be in binary equivalents, we can use any binary SOP generation and minimization technique to get $g_0$ and $g_1$. Here, for demonstration, we use Karnaugh's mapping technique (K-map), but it is possible to use other techniques such as espresso heuristic logic minimizer. Once $g_0$ and $g_1$ are obtained, the rest of the process to get quaternary SOP is same regardless of the binary SOP generation technique.

Let us separate the two constituent binary parts of the quaternary SOP as $f_1$ and $f_0$, where,
$$f_0 \equiv F.1 \qquad (67)$$
$$f_1 \equiv \sim (F.2) = \sim F.1 \qquad (68)$$

If we have $k_1 + k_3$ non-zero minterms for $f_0$ and $k_2 + k_3$ non-zero minterms for $f_1$ from two different K-maps. Therefore, we can write:

$$F_0 = F.1 \equiv f_0 = \sum \mathbf{g_0} \quad ; k_1 + k_3 \text{ terms} \qquad (69)$$

$$F_1 = \sim F.1 \equiv f_1 = \sum \mathbf{g_1} \quad ; k_2 + k_3 \text{ terms} \qquad (70)$$

where $\mathbf{g_0}$ and $\mathbf{g_1}$ are vectors of binary minterms. Once $f_0$ and $f_1$ are obtained, quaternary SOP function $F$ can be written readily.

$$F = \sim F_1.2 + F_0.1 \equiv \langle f_1, f_0 \rangle \qquad (71)$$

Eq. (71) results in a number of transformations that convert binary minterms directly into their quaternary counterparts,

which are given in Table III. Here, $X$ is any quaternary proposition and $x_0$, $x_1$ are its component bits that appear in binary SOP. Since the binary SOPs and their transformations in Table III contain only OR, AND, bitswap and NOT, we see that only these four operators are necessary and sufficient to describe any quaternary function. In Appendix A, we show how to obtain form-II of SOP for three different functions.

TABLE III : TRANSFORMATION PAIRS FOR BINARY-TO-QUATERNARY FORM-II SOP CONVERSION

| $f_0 \to F = F_0.1$ | $f_1 \to F = \sim F_1.2$ |
|---|---|
| $x_0 \equiv X.1$ | $x_0 \equiv \sim X.2$ |
| $\overline{x_0} \equiv \overline{X}.1$ | $\overline{x_0} \equiv \sim \overline{X}.2$ |
| $x_1 \equiv \sim X.1$ | $x_1 \equiv X.2$ |
| $\overline{x_1} \equiv \sim \overline{X}.1$ | $\overline{x_1} \equiv \overline{X}.2$ |

Another important feature of form-II of SOP is that it can be used to directly convert a binary function into quaternary. If a binary system has $2m$ inputs and $2n$ outputs, these $2m$ inputs can be grouped as $m$ quaternary inputs. Then we can directly convert the $2n$ binary outputs into $n$ quaternary outputs.

### C. Similarity between MIN, MAX and AND, OR operators

We can express MIN and MAX operators in form-II of SOP (please refer to Appendix A for derivation), as given below,

$$MIN(A,B) = (A.B + \sim A. \sim \overline{B}.B + \sim B. \sim \overline{A}.A).1 + (A.B).2 \tag{72}$$

$$MAX(A,B) = (\sim A.A + \sim B.B + \sim \overline{A}.B + \sim \overline{B}.A).1 + (A+B).2 \tag{73}$$

In (72) and (73), if we put any values of $A$ and $B$ except $(A,B) = (1,2)$ or $(2,1)$, then we find that,

$$MIN(A,B) = A.B \tag{74}$$

$$MAX(A,B) = A+B \tag{75}$$

where, $(A,B) \neq (1,2), (2,1)$. Now, (74) and (75) are true for any finite number of inputs as long as the set of inputs do not contain both 1 and 2 simultaneously. According to (55), in form-I of SOP, each minterm $M_D$ is the AND of several $E(\mathbf{X},\mathbf{V})$ and $D$ literals. Since $E(\mathbf{X},\mathbf{V})$ returns only 0 or 3 and the only literal that can have a value of 1 or 2 is $D$, there is no way both 1 and 2 can appear as inputs of an AND gate; thus MIN and AND are effectively equivalent for each minterm. Similarly, only one minterm remains non-zero at a time, so the OR stage may not have both 1 and 2 as inputs together, making OR and MAX functionally equivalent. Therefore, in case of form-I of SOP, AND and OR can be replaced by MIN and MAX functions, respectively. It should be noted that this argument is not valid for form-II of SOP.

There are other examples where MIN(MAX) is equivalent to AND(OR), such as the design of decoder and multiplexer that will be discussed later. This is an important feature because many of the existing quaternary logic schemes have MIN and MAX as operators and these operators have already been realized physically [16],[17].

## V. COMPUTATION OF UPPER BOUND OF GATE COUNT AND GATE DEPTH FOR SOP EXPRESSIONS

In this section we will compute the maximum number of gates and gate depth required to evaluate the SOP expressions of form-I or II for any arbitrary function. We assume the gates to be made of SWSFETs as described in [27]-[29] and base our calculations particularly on this technology. Each dyadic AND and OR gate consists of three SWSFETs in two levels, while each unary operation takes only one SWSFET. Equality, however, takes five SWSFETs spanned in two levels.

The following two lemmas are derived to compute the gate count and gate depth of a multi-level AND or OR gate array (please refer to Appendix B for derivation).

**Lemma 3**: If an AND (OR) gate may not take more than $v$ inputs, then it is possible to compute the AND (OR) of $n$ propositions using exactly $\left\lceil \frac{n-1}{v-1} \right\rceil$ gates.

**Lemma 4**: If an AND (OR) gate may not take more than $v$ inputs, then it is possible to compute the AND (OR) of $n$ propositions within $\left\lceil \log_v n \right\rceil$ depth of operators.

### A. Computation of upper bound of gate count and gate depth for form-I of SOP

For form-I of SOP, any minterm that produces a non-zero output is represented with literals consisting of equality operators. Suppose we have a function with $n$ arguments. So, the truth table of this function has $4^n$ rows. If the truth table of the function is expressed as a two dimensional map (different from K-map) as shown in Table IV for $n = 2$, the worst case is observed if no two columns or two rows are identical and the function never gives an output of 3 or 0.

TABLE IV : WORST CASE TRUTH TABLE OF FORM-I OF SOP

| A\B | 0 | 1 | 2 | 3 |
|---|---|---|---|---|
| 0 | 2 | 1 | 2 | 1 |
| 1 | 1 | 2 | 1 | 2 |

| 2 | 2 | 1 | 1 | 2 |
|---|---|---|---|---|
| 3 | 1 | 2 | 2 | 1 |

Since there are $n$ inputs and each of them may be equal to 0, 1, 2 or 3; there are at most $4n$ equality operations in parallel to calculate all these literals. Then, there are $4^n$ minterms, each containing $n$ literals and a constant value for that minterm output. Starting with these literals, each minterm takes $N_1$ gates spanned in $d_1$ gate levels as given below (from Lemma 3 and 4)-

$$d_1 = \lceil \log_{v_1}(n+1) \rceil, \quad N_1 = \lceil \frac{n}{v_1 - 1} \rceil \tag{76}$$

Here, $v_1$ is the maximum number of inputs to an AND gate. There are $4^n$ minterms calculated in parallel, each requiring $N_1$ number of gates.

The OR of these $4^n$ minterms are calculated using $N_2$ gates and spanned in $d_2$ gate levels as given below -

$$d_2 = \lceil \log_{v_2} 4^n \rceil, \quad N_2 = \lceil \frac{4^n - 1}{v_2 - 1} \rceil \tag{77}$$

Here, $v_2$ is the maximum number of inputs to an OR gate.

Now, each literal is the output of an equality operation, which is assumed to take $N_0$ number of gates spanned in $d_0$ number of gate levels. So the total number of gates is calculated as -

$$N = 4nN_0 + 4^n \lceil \frac{n}{v_1 - 1} \rceil + \lceil \frac{4^n - 1}{v_2 - 1} \rceil \tag{78}$$

The maximum gate depth is calculated as -

$$d = d_0 + \lceil \log_{v_1}(n+1) \rceil + \lceil \log_{v_2} 4^n \rceil \tag{79}$$

Considering SWSFET technology, we get the following transistor count and depth for $v_1 = v_2 = 2$,

$$N_T = 20n + 3\left(4^n(n+1) - 1\right) \tag{80}$$

$$d_T = 2 + 2\left(\lceil \log_2(n+1) \rceil + 2n\right) \tag{81}$$

### B. Computation of upper bound of gate count and gate depth for form-II of SOP

We need the literals $X, \overline{X}, \sim X$ and $\sim \overline{X}$ to write form-II of SOP for any function, where $X$ is an argument of the function. A literal along with its NOT never appear in the same minterm; and only one of $\sim X$ or $\sim \overline{X}$ can co-exist with $X$ or $\overline{X}$ in the same minterm. So we need at most two of the four propositions involving $X$ as mentioned above. Therefore, we can conclude that if a function consists of $n$ arguments, there may be at most $2n$ propositions in a single minterm. So if we consider that the maximum number of propositions of a single AND gate is $v_1$, we can conclude from Lemma 3 and 4 that the maximum number of gates and maximum gate depth for any minterm will be given by -

$$d_1 = \lceil \log_{v_1} 2n \rceil, \quad N_1 = \lceil \frac{2n - 1}{v_1 - 1} \rceil \tag{82}$$

To calculate the upper bound of gate count and gate depth, we need to consider the worst case for $f_1$ and $f_0$, each of which has the checkerboard formation of 1's and 0's like the one shown in Table V. This is definitely the worst case because if we convert any 1 to 0 or any 0 to 1, then either the number of minterms or number of propositions in a single minterm or both will reduce.

TABLE V : K-MAP FOR WORST CASE OF $F_1$ OR $F_0$

| $\langle b_1, b_0 \rangle$ \ $\langle a_1, a_0 \rangle$ | 0,0 | 0,1 | 1,1 | 1,0 |
|---|---|---|---|---|
| 0,0 | 0 | 1 | 0 | 1 |
| 0,1 | 1 | 0 | 1 | 0 |
| 1,1 | 0 | 1 | 0 | 1 |
| 1,0 | 1 | 0 | 1 | 0 |

If the worst case occurs, half of the entries of the K-map must be *1*. There are $2^{2n}$ entries in a binary K-map for $n$ quaternary propositions. Therefore, K-map for $f_1$ or $f_0$ may have at most $2^{2n-1}$ non-zero entries which is theoretically the maximum number of minterms. Now, if we limit the number of propositions of OR gate to be $v_2$, the maximum depth and maximum number of gates for the OR operation for either $F_1$ or $F_0$ are found to be-

$$d_2 = \lceil \log_{v_2} 2^{2n-1} \rceil, \quad N_2 = \left\lceil \frac{2^{2n-1} - 1}{v_2 - 1} \right\rceil \tag{83}$$

The computation of $f_1$ and $f_0$ can be done in parallel and the depths are same for both $f_1$ and $f_0$ in the worst case. If the gate depth required to calculate all literals (except $X$ itself) $\overline{X}$, $\sim X$ and $\sim \overline{X}$ is given by $d_0$ and the total depth is given by $d$, then -

$$d = d_0 + \lceil \log_{v_1} 2n \rceil + \lceil \log_{v_2} 2^{2n-1} \rceil + 2 \tag{84}$$

Here 2 is added for the computation of $\sim F_1.2 + F_0.1$, the bitswap operation shown here is distributed over the literals like the transformations in Table IV and thus this is not counted separately.

Since we need different gates for the computation of $f_1$ and $f_0$ in parallel, we need AND gates for at most $4^n$ minterms and OR gates for both $F_0$ and $F_1$. Therefore, if the number of gates required to calculate all literals except $X$ itself is given by $N_0$ and the total number of gates is given by $N$, then -

$$N = N_0 + 4^n N_1 + 2N_2 + 3 \tag{85}$$

Here, 3 is added to account for the AND and OR gates needed to compute $\sim F_1.2 + F_0.1$. We can also calculate $N_0$ as $3n$, since each argument has three literals to be calculated apart from $X$ itself. Therefore, the final expression for total number of gates is -

$$N = 3n + 4^n \left\lceil \frac{2n-1}{v_1 - 1} \right\rceil + 2\left\lceil \frac{2^{2n-1} - 1}{v_2 - 1} \right\rceil + 3 \tag{86}$$

Considering SWSFET technology, we get the following transistor count and depth for $v_1 = v_2 = 2$,

$$N_T = 3\left(2^{2n+1} n + n + 1\right) \tag{87}$$

$$d_T = 1 + 2\left(\lceil \log_2 2n \rceil + 2n + 1\right) \tag{88}$$

*C. Salient features of form-I and form-II of SOP*

Since the two forms of SOP are derived in different ways, it is often difficult to tell beforehand which form is more efficient. However, there are certain applications that favor one form over another. The design of many logic circuits that rely on a look-up table-like working principle, such as a multiplexer and any other circuit based on it, becomes very straight-forward and efficient if form-I of SOP is utilized (refer to Section VI). This is also true normally for functions with many inputs where the number of minterms is small compared to the number of variables and those terms are located sparsely in the truth table. Form-I of SOP also exists in literature in various analogous forms, having various design and implementation techniques already been developed for it [17].

However, the biggest advantage of form-II of SOP is the flexibility it brings in designing large and complex logic functions. Any binary logic simplification technique can be used with it and it allows direct transformation of binary logic circuits in quaternary. Also, form-II of SOP is subject to more optimization techniques than form-I as discussed in [25]. Not only that, methodological design of tree-based logic circuits such as fast adders, comparators and encoders becomes much simpler of a problem if we start with form-II of SOP [21],[22],[26]. Finally, if a large binary system is to be converted into a quaternary system by replacing the internal circuitry of the system, while putting binary-to-quaternary encoders and quaternary-to-binary decoders on input and output sides of the system respectively, then form-II of SOP would more likely be a reasonable choice due to its close relation with binary SOP expressions.

## VI. DESIGN OF SOME IMPORTANT COMBINATIONAL QUATERNARY CIRCUITS

Combinational circuits are the vital elements for any digital system. Our proposed quaternary operators can be employed efficiently to design many common combinational circuits [20]-[26]. We generalize some of them here using the form-II of SOP.

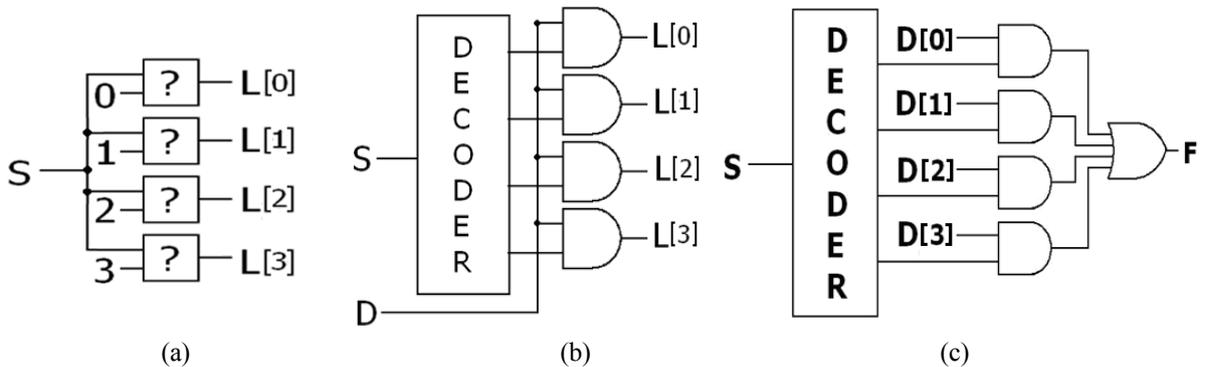

Fig. 2. Quaternary 4-line (a) decoder, (b) demultiplexer and (c) multiplexer.

A quaternary 1-to-4 decoder (Fig. 2a, Table VI) has one input $S$ and four outputs, defined by the array $\boldsymbol{L}$. Only one of the outputs can be equal to 3 at a time, all other outputs remain 0. The outputs are defined by the following equation -

$$\mathbf{L}[i] = S^i \quad ; \quad i = 0, 1, 2, 3 \tag{89}$$

In general, for $n$-to-$4^n$ line decoder, we have $n$ input lines given by the array, $\mathbf{S} = \{S_1, S_2, ...., S_n\}$. The array $V$ denotes each of $4^n$ possible combinations of inputs. The state of any output line is denoted by-

$$\mathbf{L}[j] = \prod \mathbf{E}(\mathbf{S}, \mathbf{V_j}) \tag{90}$$

A 1-to-4 demultiplexer (Fig. 2b, Table VI) is same as 1-to-4 decoder, but the output passes an additional data input $D$ through one of the output lines set by the selector input $S$. The outputs of the 1-to-4 and $n$-to-$4^n$ demultiplexers are expressed by the following equations -

$$\mathbf{L}[i] = D \cdot S^i \quad ; i = 0, 1, 2, 3 \tag{91}$$

$$\mathbf{L}[j] = D \cdot \prod \mathbf{E}(\mathbf{S}, \mathbf{V_j}) \tag{92}$$

TABLE VI : TRUTH TABLES OF QUATERNARY 1-TO-4 DECODER, DEMULTIPLEXER AND 4-TO-1 MULTIPLEXER

| $S$ | Decoder output ($L$) | | | | Demultiplexer output ($L$) | | | | Multiplexer output (F) |
|---|---|---|---|---|---|---|---|---|---|
| | [0] | [1] | [2] | [3] | [0] | [1] | [2] | [3] | |
| 0 | 3 | 0 | 0 | 0 | $D$ | 0 | 0 | 0 | $D_0$ |
| 1 | 0 | 3 | 0 | 0 | 0 | $D$ | 0 | 0 | $D_1$ |
| 2 | 0 | 0 | 3 | 0 | 0 | 0 | $D$ | 0 | $D_2$ |
| 3 | 0 | 0 | 0 | 3 | 0 | 0 | 0 | $D$ | $D_3$ |

A multiplexer can be constructed using a decoder. There are as many data inputs as the outputs of the decoder and each output is a minterm involving the decoder output and the corresponding data input (product of decoder output $\mathbf{L}$ and data input $\mathbf{D}$). The OR of all such minterms give the output of the multiplexer, expressed as a SOP of form-II. If there are $n$ selectors and $4^n$ data inputs, then the multiplexer output is given by -

$$M = \sum_{j=0}^{4^n-1} (\mathbf{D}[j] \cdot \mathbf{L}[j]) = \sum_{j=0}^{4^n-1} \left( \mathbf{D}[j] \cdot \prod \mathbf{E}(\mathbf{S}, \mathbf{V_j}) \right) \tag{93}$$

For $n = 1$ (Fig. 2c, Table VI),

$$M = \sum \left( \mathbf{D}[i] \cdot S^i \right) \quad ; \quad i = 0, 1, 2, 3 \tag{94}$$

## VII. COMPARISON WITH EXISTING LOGIC SYSTEMS

In this section we present the comparison between the proposed algebra and some other existing variants of quaternary logic. Here we would like to make the comparison in both gate level and transistor-level, but direct transistor level comparison is not possible because no other version of quaternary logic has been implemented using SWSFET, a technology fully compatible with the proposed logic. Gate/transistor count, however, is a relative measure of complexity associated with the design of logic circuits. Since a comparison with SWSFET devices with similar binary FETs is presented in [27], we will present a transistor-level comparison in our comparison with binary logic system based on that.

TABLE VII : COMPARISON OF THE PROPOSED LOGIC WITH OTHER QUATERNARY VARIANTS

| Device | Gate Count (Literature) | Gate/SWSFET Count (Current Work) |
|---|---|---|
| 2-to-16 decoder | 16x 3-digit QCSGs, 20x QSGs [9] equivalent: 80x 2-digit QCSGs, 20x QSGs | 8x unary equalities, 16x 2-input AND equivalent: 8 + 48 (= 56) SWSFETs |
| 2-to-16 decoder | 16x QSGs, 16x 3-digit modified M-S [12] equivalent: 80x QSGs, 80x M-S | 8x unary equalities, 16x 2-input AND equivalent: 8 + 48 (= 56) SWSFETs |
| 2-to-16 demux | 16x 3-digit QCSGs, 20x QSGs, 16x 3-input Toffoli [9] equivalent: 80x 2-digit QCSGs, 244x QSGs, 208x M-S | 8x unary equalities, 32x 2-input AND equivalent: 8 + 96 (= 104) SWSFETs |
| 16-to-1 mux | 16x 3-digit QCSGs, 20x QSGs, 16x 3-input Toffoli [9] equivalent: 80x 2-digit QCSGs, 244x QSGs, 208x M-S | 8x unary equalities, 32x 2-input AND, 15x 2-input OR equivalent: 8 + 96 + 45 (= 149) SWSFETs |
| 1-to-4 demux | 8x 2-digit M-S, 4x QSGs, 4x 3-input Toffoli [12] equivalent: 60x QSGs, 60x 2-digit M-S | 4x unary equalities, 4x 2-input AND equivalent: 4 + 12 (= 16) SWSFETs |
| 4-to-1 mux | 8x 2-digit M-S, 4x QSGs, 4x 3-input Toffoli [12] equivalent: 60x QSGs, 60x 2-digit M-S | 4x unary equalities, 4x 2-input AND, 3x 2-input OR equivalent: 4 + 12 + 9 (= 25) SWSFETs |

Khan et al. used quaternary shift gates (QSG) and 2-digit quaternary controlled shift gates (QCSG) or Muthukrishnan-Stroud (M-S) gates to show the realization of several higher level gates like Feynman gates and Toffoli gates

[9],[12],[32],[33]. Using these gates, quaternary decoder, multiplexer and demultiplexer were designed. In the context of quantum logic, qudit means quantum digit. To implement 2-to-16 decoder with active-1 output, Khan [9] used 16 of 3-digit QCSGs and 20 of QSGs. Since each 3-digit QCSG required 5 of 2-digit QCSGs, this design took 80 of 2-digit QCSGs and 20 of QSGs. For both 16-to-1 multiplexer and 2-to-16 demultiplexer, they used 16 of 3-input Toffoli gates in addition to their 2-to-16 decoder circuit proposed in the same work.

Another design of 2-to-16 decoder circuit was given by Khan [12], where QSGs and 3-qudit modified M-S gates were used, 16 of each type. Each modified M-S gate individually required 5 of 2-qudit M-S gates and 4 QSGs. 1-to-4 demultiplexer and 4-to-1 multiplexer were also designed, each with 4 of 3-input Toffoli gates, 8 M-S gates and 4 QSGs. A 3-input Toffoli gate was realized using 13 of M-S gates and 14 of QSGs.

The above results from the literature are listed in Table VII along with the gate/SWSFET counts for the same devices using the proposed logic. Although a direct comparison is not possible in terms of the gate count, significant reduction in design complexity is readily evident from the results of the present work.

In Table VIII, we compare some binary logic blocks with their quaternary counterparts. Here binary devices are assumed to be implemented using CMOS gates which share the same or similar technology as the SWSFETs. NAND/NOR gates are restricted to 2-inputs only and number of inputs/outputs are taken to be an exponent of 2, to avoid redundancy in quaternary implementation. This is why demultiplexer and multiplexer are assumed to have dual data lines to match the quaternary equivalent. For quaternary implementation, it is assumed that all inputs and outputs are available in quaternary so that no encoder/decoder is necessary. It was claimed in [27]-[29] that up to 75% reduction in transistor count and up to 50% reduction in data interconnect densities could be achieved with reduced power dissipation and gate delay, if quaternary logic with SWSFET technology could be used instead of CMOS binary logic. Table VIII is in good agreement with the claim about the transistor count, as the reduction in transistor count is observed in between 44% and 80%.

TABLE VIII : COMPARISON OF THE PROPOSED LOGIC WITH BINARY LOGIC

| Binary Device | Gate Count in Binary Logic | Quaternary Gate/SWSFET Count in Current Work |
|---|---|---|
| 2-to-4 decoder | 4x NOR, 2x NOT<br>CMOS equivalent: 16 + 4 (= 20) FETs | 4x unary equalities<br>equivalent: 4 SWSFETs |
| Dual 2-to-4 demux | 4x NAND, 8x NOR, 3x NOT<br>CMOS equivalent: 16 + 32 + 6 (= 54) FETs | 4x unary equalities, 4x 2-input AND<br>equivalent: 4 + 12 (= 16) SWSFETs |
| 4-to-16 decoder | 8x NAND, 16x NOR, 4x NOT<br>CMOS equivalent: 32 + 64 + 4 (= 100) FETs | 8x unary equalities, 16x 2-input AND<br>equivalent: 8 + 48 (= 56) SWSFETs |
| Dual 4-to-1 mux | 8x NAND, 10x NOR, 4x NOT<br>CMOS equivalent: 32 + 40 + 8 (= 80) FETs | 4x unary equalities, 4x 2-input AND, 3x 2-input OR<br>equivalent: 4 + 12 + 9 (= 25) SWSFETs |

## VIII. CONCLUSION

In this paper, we have presented a novel quaternary algebra which serves as a bridge between the well-developed binary logic and the emerging quaternary logic. The algebra aids in transforming any binary function into its quaternary version and allows the quaternary functions to be simplified and manipulated using the simplifying techniques of the binary logic. It also includes operators that are commonly found in other existing quaternary logic variants, and thus is capable of handling logical expressions derived in other existing quaternary logic systems. Besides, using the unique properties of the operators defined in this logic, we have established two methods to express any quaternary function as a sum-of-products (SOP) expression, one of them being completely unique to the proposed logic. We have presented the theoretical analyses of both forms of SOPs and discussed their unique applications. Finally several simple logic circuits have been presented and based on them, a comparative study of the proposed logic with binary and other quaternary logic systems have been made.

# Appendix A

*Logical synthesis of equality operator*

The truth table of $E(A,B)$ can be written in the matrix form as shown in Table A-I.

TABLE A-I : TRUTH TABLE OF $E(A,B)$

| A \ B | 0 | 1 | 2 | 3 |
|---|---|---|---|---|
| 0 | 3 | 0 | 0 | 0 |
| 1 | 0 | 3 | 0 | 0 |
| 2 | 0 | 0 | 3 | 0 |
| 3 | 0 | 0 | 0 | 3 |

The binary truth tables for $e_0(a_0,a_1,b_0,b_1)$ and $e_1(a_0,a_1,b_0,b_1)$ are identical, having *1*'s at the locations where $E(A,B)$ has *3*'s.

The binary functions are thus identical as well and can be written as –

$$e_0(a_0,a_1,b_0,b_1) = e_1(a_0,a_1,b_0,b_1) = \overline{a_0}.\overline{a_1}.\overline{b_0}.\overline{b_1} + \overline{a_0}.a_1.\overline{b_0}.b_1 + a_0.\overline{a_1}.b_0.\overline{b_1} + a_0.a_1.b_0.b_1 \quad \text{(A-1)}$$

From (A-1), we calculate $E(A,B)$ -

$$E(A,B) = \sim\overline{A}.\overline{A}.\sim\overline{B}.\overline{B} + \sim\overline{A}.A.\sim\overline{B}.B + \sim A.\overline{A}.\sim B.\overline{B} + \sim A.A.\sim B.B \quad \text{(A-2)}$$

Now, we can manipulate $E(A,B)$ further to get a more compact expression.

$$E(A,B) = \sim\overline{A}.\sim\overline{B}.(A.B + \overline{A}.\overline{B}) + \sim A.\sim B.(A.B + \overline{A}.\overline{B}) = (A.B + \overline{A}.\overline{B}).(\sim\overline{A}.\sim\overline{B} + \sim A.\sim B) \quad \text{(A-3)}$$

Thus, we get two equivalent expressions -

$$E(A,B) = \overline{(A \oplus B)} \cdot \sim(A \oplus B) \quad \text{(A-4)}$$

$$E(A,B) = \overline{(A \oplus B) + \sim(A \oplus B)} \quad \text{(A-5)}$$

Here, instead of using just AND, OR, NOT and bitswap, we used XOR/XNOR operators. In this way, the functional operators can be used to reduce SOP expressions of form-II.

*Logical synthesis of MIN and MAX operator*

We will express two functional operators $MIN(A, B)$ and $MAX(A, B)$ in form-II of SOP where $A$ and $B$ are two quaternary propositions. Let us first implement $MIN(A, B)$. The truth table of the function is shown in Table A-II.

TABLE A-II : TRUTH TABLE OF QUATERNARY MIN FUNCTION

| B \ A | 0 | 1 | 3 | 2 |
|---|---|---|---|---|
| 0 | 0 | 0 | 0 | 0 |
| 1 | 0 | 1 | 1 | 1 |
| 3 | 0 | 1 | 3 | 2 |
| 2 | 0 | 1 | 2 | 2 |

Now, we consider the cells which contain *1* or *3* as true for $min_0$. Similarly, we consider the cells which contain *2* or *3* as true for $min_1$. The remaining cells will contain *0*. The K-map for $min_0$ is given in Table A-III.

TABLE A-III : K-MAP FOR $MIN_0$ FUNCTION

| $\langle b_1,b_0\rangle$ \ $\langle a_1,a_0\rangle$ | 0,0 | 0,1 | 1,1 | 1,0 |
|---|---|---|---|---|
| 0,0 | 0 | 0 | 0 | 0 |
| 0,1 | 0 | 1 | 1 | 1 |
| 1,1 | 0 | 1 | 1 | 0 |
| 1,0 | 0 | 1 | 0 | 0 |

From the K-map given above, we can write the following expressions:

$$min_0(a_1,a_0,b_1,b_0) = a_0.b_0 + a_1.\overline{b_1}.b_0 + b_1.\overline{a_1}.a_0 \quad \text{(A-6)}$$

$$MIN_0(A,B) = (A.B + \sim A. \sim \overline{B}.B + \sim B. \sim \overline{A}.A).1 \quad \text{(A-7)}$$

Now, the K-map for $min_1$ is given in Table A-IV.

TABLE A-IV : K-MAP FOR $MIN_1$ FUNCTION

| $\langle b_1,b_0\rangle$ \ $\langle a_1,a_0\rangle$ | 0,0 | 0,1 | 1,1 | 1,0 |
|---|---|---|---|---|
| 0,0 | 0 | 0 | 0 | 0 |
| 0,1 | 0 | 0 | 0 | 0 |
| 1,1 | 0 | 0 | 1 | 1 |
| 1,0 | 0 | 0 | 1 | 1 |

From the K-map we can get the following expression:
$$min_1(a_1,a_0,b_1,b_0) = a_1.b_1 \tag{A-8}$$
$$MIN_1(A,B) = (A.B).2 \tag{A-9}$$
Now, we can combine (A-7) and (A-9) to get the complete function:
$$MIN(A,B) = (A.B + \sim A.\overline{B}.B + \sim B.\overline{A}.A).1 + (A.B).2 \tag{A-10}$$
Similarly, we can derive $MAX_1(A, B)$ and $MAX_0(A, B)$ from Table A-V.

TABLE A-V : TRUTH TABLE OF QUATERNARY MAX FUNCTION

| B \ A | 0 | 1 | 3 | 2 |
|---|---|---|---|---|
| 0 | 0 | 1 | 3 | 2 |
| 1 | 1 | 1 | 3 | 2 |
| 3 | 3 | 3 | 3 | 3 |
| 2 | 2 | 2 | 3 | 2 |

Using the same technique described for $MIN(A, B)$, we can write the following expressions:
$$max_0(a_1,a_0,b_1,b_0) = a_1.a_0 + b_1.b_0 + \overline{a_1}.b_0 + \overline{b_1}.a_0 \tag{A-11}$$
$$MAX_0(A,B) = (\sim A.A + \sim B.B + \sim \overline{A}.B + \sim \overline{B}.A).1 \tag{A-12}$$
$$max_1(a_1,a_0,b_1,b_0) = a_1 + b_1 \tag{A-13}$$
$$MAX_1(A,B) = (A+B).2 \tag{A-14}$$
Now, we can combine (A-12) and (A-14) to get the complete function:
$$MAX(A,B) = (\sim A.A + \sim B.B + \sim \overline{A}.B + \sim \overline{B}.A).1 + (A+B).2 \tag{A-15}$$

## **Appendix B**

**Lemma 3**: If an AND (OR) gate may not take more than $v$ inputs, then it is possible to compute the AND (OR) of $n$ propositions using exactly $\left\lceil \frac{n-1}{v-1} \right\rceil$ gates.

**Proof:** Suppose, there are $n$ propositions and we want to calculate the AND of them, where the maximum number of inputs to an AND gate is $v$. If $n > v$, we will need more than one gates to perform the computation. Now, the number of gates is minimized if at most one gate is allowed to have less than $v$ inputs. To minimize delay, the inputs are divided into groups and processed in parallel AND gates. If there are $x_0$ number of gates each processing exactly $v$ number of such propositions, then -
$$v\,x_0 + r_0 = n \tag{B-1}$$
where,
$$x_0 = \left\lfloor \frac{n}{v} \right\rfloor \quad , \quad r_0 = n - v\left\lfloor \frac{n}{v} \right\rfloor \tag{B-2}$$
These $x_0$ gates have $x_0$ outputs and also there are $r_0$ propositions left if $n$ is not absolutely divisible by $v$. So at the second level, there are $x_0 + r_0$ propositions. If this level has $x_1$ gates and $r_1$ remaining propositions, then -
$$x_0 = v\,x_1 + r_1 - r_0 \tag{B-3}$$
Generalizing for higher levels, we get the following for the $m$-th level,
$$x_{m-1} = v\,x_m + r_m - r_{m-1} \tag{B-4}$$
Now, if $(m+1)$-th level is the last level, then it must have only one gate so that the single output gives the AND of all $n$ propositions. Let us assume that the last gate may have fewer than $v$ propositions and the number of unused inputs is $\delta < v$. Therefore we get the following equation-
$$x_m = v - \delta - r_m \tag{B-5}$$
Summing the equations for all the stages between $x_0$ and $x_m$ using (B-3) and (B-5), we get the following equation -
$$\sum_{i=0}^{m} x_i = v \sum_{i=1}^{m} x_i + v - \delta - r_0 \tag{B-6}$$

This can be simplified to the following form-
$$\sum_{i=1}^{m} x_i = \frac{x_0 - v + \delta + r_0}{v-1} = \frac{\left\lfloor \frac{n}{v} \right\rfloor - v + \delta + n - v\left\lfloor \frac{n}{v} \right\rfloor}{v-1} \tag{B-7}$$
So, the total number of gates is -

$$N = 1 + x_0 + \sum_{i=1}^{m} x_i = \frac{v\left\lfloor\frac{n}{v}\right\rfloor + \delta + n - v\left\lfloor\frac{n}{v}\right\rfloor - 1}{v - 1} \tag{B-8}$$

After simplification,

$$N = \frac{n-1}{v-1} + \frac{\delta}{v-1} \tag{B-9}$$

The quantity $\delta$ is used to account for the unused inputs to the final gate, so that $N$ becomes an integer. Therefore, $N$ can be expressed as -

$$N = \left\lceil \frac{n-1}{v-1} \right\rceil \tag{B-10}$$

Although we assumed so far that the $\delta$ unused inputs might only exist at the final stage, but it was done only for the sake of simplicity. Even if other stages had unused input capacity, (B-10) would still hold, given that the total number of such unused inputs did not differ from $\delta$.

**Lemma 4**: If an AND (OR) gate may not take more than $v$ inputs, then it is possible to compute the AND (OR) of $n$ propositions within $\lceil \log_v n \rceil$ depth of operators.

**Proof:** The proof continues from the proof of Lemma 3. If $n = v^k$ for any positive integer $k$, then after each level, the number of propositions to be calculated at the next level is reduced by a factor of $v$. Now, if gate depth is denoted by $d$, we get the following result for $n = v^k$,

$$d = k \tag{B-11}$$

If $n = v^{k+1}$, then we have -

$$d = k + 1 \tag{B-12}$$

The results for $n = v^k$ and $n = v^{k+1}$ can be summarized for gate depth $d$ by the following equation-

$$d = \log_v n \tag{B-13}$$

Now, if $d$ lies between $k$ and $k+1$, i.e. $v^k < n < v^{k+1}$, then we can express $n$ as a polynomial of $v$ as given below -

$$n = a_k v^k + a_{k-1} v^{k-1} + \ldots + a_0 \tag{B-14}$$

where $a_i \in \{0, 1, 2, \ldots, v-1\}$ and at least two of all $a_i$ including $a_k$ must be non-zero.

This means the $n$ propositions are grouped in several sub-groups so that the number of propositions in each sub-group is equal to a non-zero term of (B-14); these sub-groups are processed in parallel. The group having lower number of propositions (corresponding to a lower order term) is processed in fewer stages and the final output is then merged with a larger group. Since there are $v - a_i \geq 1$ unused inputs at any $(i+1)$-th level, this single proposition does not increase the number of gate levels associated with that term.

This argument holds for all terms up to the highest order term and it is always possible to calculate the AND of $n$ propositions in (B-14) to be calculated within $k+1$ gate levels. Which is also true for $n = v^{k+1}$.

Considering the three cases of $n = v^k$, $n = v^{k+1}$ and $v^k < n < v^{k+1}$ for arbitrary $v$ and $k$, we conclude that the gate depth $d$ can be given by the following formula -

$$d = \lceil \log_v n \rceil \tag{B-15}$$